\documentclass[twocolumn]{aastex62}

\newcommand{\kms}{km~s$^{-1}$}

\newcommand{\lsun}{$L_\odot$}

\newcommand{\degrees}{$^{\circ}$}

\received{October 16, 2018}
\revised{November 9, 2018}
\accepted{November 9, 2018}
\submitjournal{ApJL}

\shorttitle{A FRAGMENTED KEPLERIAN DISK AROUND A PROTO-O STAR}
\shortauthors{Ilee et al.}

\begin{document}

\title{G11.92$-$0.61 MM\,1: A FRAGMENTED KEPLERIAN DISK SURROUNDING A PROTO-O STAR}

\correspondingauthor{John D. Ilee}
\email{j.d.ilee@leeds.ac.uk}

\author[0000-0003-1008-1142]{J.\ D.\ Ilee}
\affil{School of Physics \& Astronomy, University of Leeds, Leeds LS2 9JT, UK}
\affil{Institute of Astronomy, University of Cambridge, Cambridge CB3 0HA, UK}

\author[0000-0001-6725-1734]{C.\ J.\ Cyganowski}
\affiliation{SUPA, School of Physics and Astronomy, University of St Andrews, North Haugh, St Andrews KY16 9SS, UK}

\author[0000-0002-6558-7653]{C.\ L.\ Brogan}
\affiliation{NRAO, 520 Edgemont Rd, Charlottesville, VA 22903, USA}

\author[0000-0001-6492-0090]{T.\ R.\ Hunter}
\affiliation{NRAO, 520 Edgemont Rd, Charlottesville, VA 22903, USA}

\author[0000-0003-1175-4388]{D.\ H.\ Forgan}
\affiliation{SUPA, School of Physics and Astronomy, University of St Andrews, North Haugh, St Andrews KY16 9SS, UK}
\affiliation{St Andrews Centre for Exoplanet Science, University of St Andrews, St Andrews KY16 9SS, UK}

\author[0000-0002-9593-7618]{T.\ J.\ Haworth}
\affiliation{Imperial College London, Blackett Laboratory, Prince Consort Road, London SW7 2AZ, UK}

\author[0000-0003-4288-0248]{C.\ J.\ Clarke}
\affil{Institute of Astronomy, University of Cambridge, Cambridge CB3 0HA, UK}

\author[0000-0001-8228-9503]{T.\ J.\ Harries}
\affiliation{University of Exeter, Stocker Road, Exeter EX4 4QL}



\begin{abstract}

We present high resolution ($\sim$300\,au) Atacama Large Millimeter/submillimeter Array (ALMA) observations of the massive young stellar object G11.92$-$0.61 MM\,1.  We resolve the immediate circumstellar environment of MM\,1 in 1.3\,mm continuum emission and CH$_{3}$CN emission for the first time.  The object divides into two main sources --- MM\,1a, which is the source of a bipolar molecular outflow, and MM\,1b, located $0\farcs57$ (1920\,au) to the South-East.  The main component of MM\,1a is an elongated continuum structure, perpendicular to the bipolar outflow, with a size of $0\farcs141\times0\farcs050$ ($480\times170$\,au).  The gas kinematics toward MM\,1a probed via CH$_{3}$CN trace a variety of scales.  The lower energy $J=12$--11 $K=3$ line traces extended, rotating gas within the outflow cavity, while the $v$8=1 line shows a clearly-resolved Keplerian rotation signature.  Analysis of the gas kinematics and dust emission shows that the total enclosed mass in MM\,1a is $40\pm5$\,M$_{\odot}$ (where between 2.2--5.8\,M$_{\odot}$ is attributed to the disk), while MM\,1b is $<0.6$\,M$_{\odot}$.  The extreme mass ratio and orbital properties of MM\,1a and MM\,1b suggest that MM\,1b is one of the first observed examples of the formation of a binary star via disk fragmentation around a massive young (proto)star.

\end{abstract}

\keywords{accretion, accretion disks --- ISM: individual objects (G11.92$-$0.61) --- stars: formation --- stars: protostars --- submillimeter: ISM}


\section{Introduction} 
\label{sec:intro}

The formation mechanisms of massive young stellar objects (MYSOs, $M_{\star}  >  8\,M_{\odot}$) are poorly understood due to their large distances and extreme embedded nature.  Models have suggested that channelling material through a  circumstellar accretion  disk can overcome the powerful feedback from the central protostar \citep[][]{krumholz_2009, kuiper_2011, rosen_2016}.  Such models predict that these disks possess significant sub-structure, including large scale spiral arms and bound fragments \citep{klassen_2016, harries_2017, meyer_2018}.  Observationally, however, it is not clear whether Keplerian circumstellar disks surround MYSOs of all masses and evolutionary stages \citep[see][for a review]{beltran_2016}, though convincing candidates are beginning to emerge \citep{johnston_2015, ilee_2016}.  In many cases, complex velocity structures, high continuum optical depths, and potential multiplicity \citep[e.g.][]{maud_2017, cesaroni_2017, beuther_2018, csengeri_2018, ahmadi_2018} make  comprehensive characterisation of the physical properties of these disks challenging.   

\smallskip

Such characterisation is important in order to connect the processes of massive star formation with the population of massive O- and B-type stars observed in the field.  High-resolution radial velocity surveys have found that $>80$ per cent of OB stars are found in close binary systems \citep[][]{chini_2012}.  Do these high-mass multiple stellar systems form via the large-scale fragmentation of turbulent cloud cores \citep[e.g.][]{fisher_2004}, or via smaller-scale fragmentation of a massive protostellar disk \citep[e.g.][]{adams_1989}?  Answering such a question requires high angular resolution observations of individual, deeply-embedded massive protostellar systems that are still in the process of formation.  

\smallskip

G11.92--0.61  MM\,1 (hereafter  MM\,1) was identified  during studies of GLIMPSE Extended Green Objects \citep[EGOs;][]{cyganowski_2008}, and is located in an infrared dark cloud (IRDC) $\sim$1\arcmin\/ SW of the more evolved massive star-forming region \emph{IRAS} 18110--1854. The total luminosity of G11.92$-$0.61 is $\sim$10$^{4}$\lsun \citep{cyganowski_2011sma,moscadelli_2016}, and its distance is 3.37$^{+0.39}_{-0.32}$\,kpc \citep[based on maser parallaxes;][]{sato_2014}.  MM\,1 drives a single, dominant bipolar molecular outflow traced by well-collimated, high-velocity $^{12}$CO(2--1) and HCO$^{+}$(1--0) emission \citep{cyganowski_2011sma}, and is coincident with a 6.7\,GHz Class II CH$_{3}$OH and strong H$_{2}$O masers \citep{hofner_1996,cyganowski_2009,breen_2011,sato_2014,moscadelli_2016}.  All of these characteristics suggest the presence of a massive (proto)star. 

\smallskip

In \citet{ilee_2016}, we analysed the properties of the centimeter and millimeter emission from MM\,1.  Our 1.3\,mm Submillimeter Array (SMA) observations (resolution $\sim$ $0\farcs46$, 1550\,au) showed consistent velocity gradients across multiple hot-core-tracing molecules oriented perpendicular to the bipolar molecular outflow.  The kinematics of these lines suggested an infalling Keplerian disk with a radius of 1200\,au, surrounding an enclosed mass of $\sim$30--60\,M$_{\odot}$, of which 2--3\,M$_{\odot}$  could be attributed to the disk.  Such a massive, extended Keplerian disk brings into question its stability against gravitational fragmentation.  In \citet{forgan_2016}, we performed a detailed analysis of MM\,1 (and other systems) utilising semi-analytic models of self-gravitating disks.  For the properties determined from our SMA observations, the disk around MM\,1 satisfies all conditions for fragmentation, with the models predicting fragment masses of $\sim$0.4\,M$_{\odot}$ for disk radii $\sim$1200\,au when accretion rates are $\gtrsim 10^{-4}$ M$_{\odot}$\,yr$^{-1}$.

\smallskip

In this Letter, we report high spatial and spectral resolution line and continuum ALMA observations of G11.92$-$0.61 MM\,1 that were designed to further characterise the circumstellar environment of this massive young stellar object, and search for evidence of disk fragmentation.  

\bigskip

\section{Observations}
\label{sec:obs}

Our ALMA observations were taken on 2017 Aug 07 (project ID 2016.1.01147.S, PI: J.~D.~Ilee) in configuration C40-7 with 46 antennas.  The projected baselines ranged from $\sim$15--2800\,k$\lambda$.  We observed in Band 6 (230\,GHz, 1.3\,mm) with four SPWs (220.26--220.73, 221.00--221.94, 235.28--236.22 and 238.35--239.29\,GHz) for an on-source time of 93\,mins. Imaging with Briggs weighting with a robust parameter of 0 yielded a synthesised beamsize of $0\farcs106\times0\farcs079$ ($360\times270$\,au), PA$=-63.7^{\circ}$ East of North, and a largest recoverable scale of $0\farcs58$ (1955\,au). Calibration, imaging and analysis were performed with CASA version 5.1.1 \citep{mcmullin_2007}. The continuum data were self-calibrated iteratively, with phase and amplitude solution times of 6 and 54 seconds, respectively, with a resulting S/N of 569 (an improvement factor of 1.4). The continuum self-calibration solutions were also applied to the line data. Continuum subtraction was performed following the method of \citet{brogan_2018}, resulting in a continuum bandwidth of 0.38\,GHz and sensitivity of 0.05\,mJy\,beam$^{-1}$.  The line  data were re-sampled to a common velocity resolution of 0.7\,km\,s$^{-1}$  to improve signal-to-noise, achieving a typical per-channel sensitivity of 1.2\,mJy\,beam$^{-1}$.

\begin{figure*}
\centering
\includegraphics[width=1.0\textwidth]{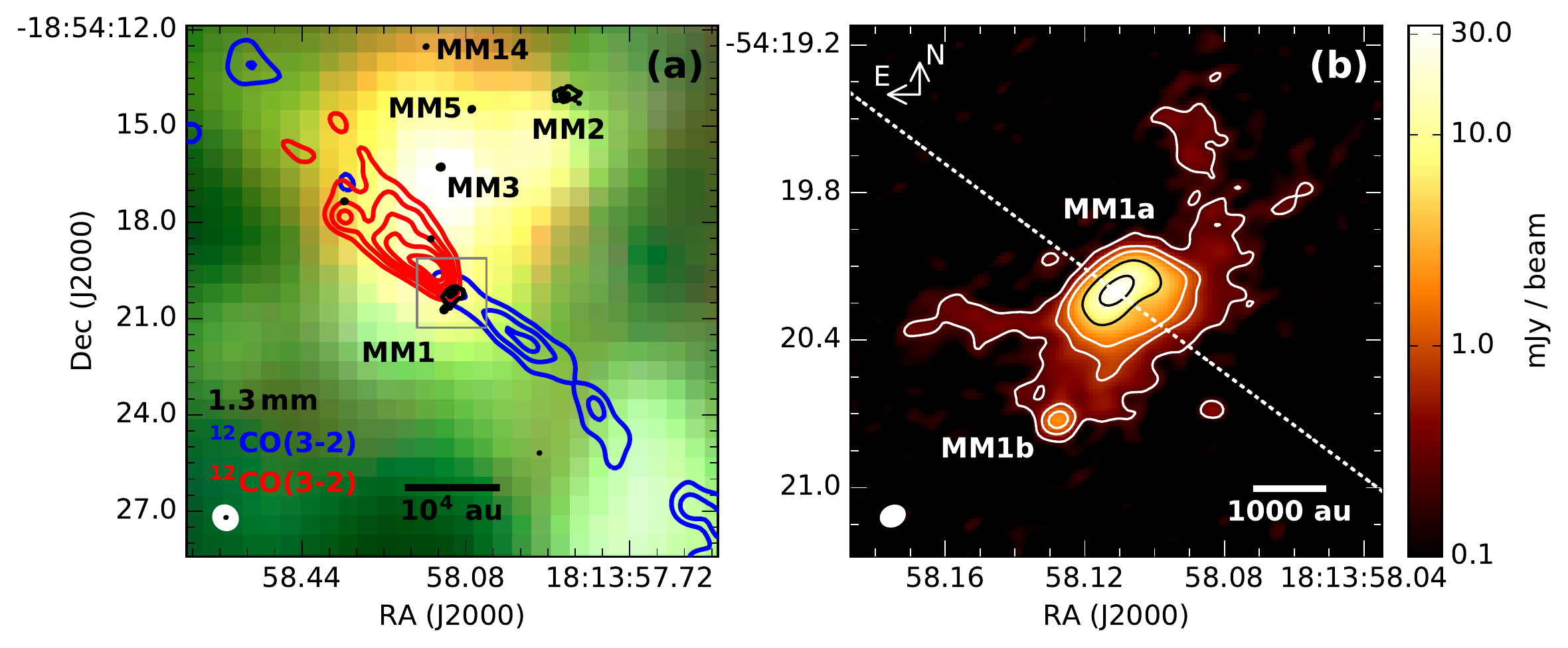}
\caption{\textbf{(a):} G11.92$-$0.61: ALMA 1.3\,mm  continuum emission (black contours) and SMA blue/redshifted $^{12}$CO(3--2) integrated intensity (blue: $-16$--$20$\,km\,s$^{-1}$, red: $50$--$74$\,km\,s$^{-1}$) overlaid on a three-color \emph{Spitzer} image (RGB: 8.0, 4.5, 3.6\,$\mu$m).  Levels:  1.3\,mm: (5,150)$\sigma$,  $\sigma=0.05$\,mJy\,beam$^{-1}$; $^{12}$CO: 0.8\,Jy\,beam$^{-1}$\,km\,s$^{-1}$ $\times$(5,10,15) (blue), $\times$(5,10,15,20,25) (red).
\textbf{(b):} Zoom view of the ALMA 1.3\,mm continuum emission towards MM\,1 (colorscale \& contours; the grey box in (a) shows the FOV of (b)).  The dotted white line shows the position angle of the $^{12}$CO outflow ($53\degr$).   Levels: (5,15,30,100,400)$\sigma$,   $\sigma=0.05$\,mJy\,beam$^{-1}$. Beams are shown at lower left.}
\label{fig:continuum}
\end{figure*}

\begin{deluxetable*}{lcccccc}
\tablewidth{0pc}
\tablecaption{Fitted properties: 1.3\,mm continuum }
\label{tab:cont}
\tablecolumns{7}
\tablehead{
\colhead{Source} & \multicolumn{2}{c}{Fitted Position (J2000)} & \multicolumn{1}{c}{Integ. Flux Density\tablenotemark{a}} & \colhead{Peak Intensity\tablenotemark{a}}          & \colhead{T$_{\rm b}$\tablenotemark{b}}    &\colhead{FWHM of deconvolved Gaussian model\tablenotemark{a}} \\
                 & \colhead{$\alpha~({^{\rm h}~^{\rm m}~^{\rm s}}$)} & \colhead{$\delta~(^\circ~\arcmin~\arcsec)$}             &  \colhead{ (mJy)}                                  & \colhead{(mJy\,beam$^{-1}$)} & \colhead{(K)}    &\colhead{($'' \times ''$ [P.A.($\arcdeg$)]) }}
\startdata
MM\,1a\tablenotemark{c}                            &                &                &                       &                                                     &        &                      \\
\phantom{xx}\phantom{$ii$}$(i)$ -- Main disk      & 18:13:58.111  & $-$18:54:20.205  & 53.2 (0.6)         &  26.8 (0.2)                           & 93       & $0.141\times0.050$ ($0.002$) [+129.4 (0.1)]    \\
\phantom{xx}\phantom{$i$}$(ii)$ -- SW excess     & 18:13:58.108  & $-$18:54:20.266  & 44.1 (2.0)         &  3.5 (0.2)                             & 11        & $0.39\times0.24$ ($0.02$) [+119 (4)]     \\
\phantom{xx}\phantom{}$(iii)$ -- W excess    & 18:13:58.104  & $-$18:54:20.140  & 10.2 (0.7)          &  4.4 (0.2)                                & 62\        & $0.17\times0.02$ ($0.02$) [+62 (3)]        \\
\hspace{1.2em}{}$(iv)$ -- Free-free\tablenotemark{d}     & 18:13:58.111  & $-$18:54:20.185  & 4.0 & 4.0                                           & ...        & ... \\                  
\vspace{-1em}\\
MM\,1b                            & 18:13:58.128  & $-$18:54:20.721 & 2.5 (0.2)            &  2.1 (0.1)                                           & 6      & $0.069 \times 0.016$ ($0.02$) [+35 (43)]     \\ 
\enddata
\tablenotetext{a}{Uncertainties are given in parentheses; for size, the listed value is the larger of the uncertainties for the two axes.}
\tablenotetext{b}{Calculated from: ($i$) the integrated flux density and the solid angle of a top-hat disk model that produces the same observed size as the Gaussian model, ($ii$) \& ($iii$) the integrated flux density and the solid angle  corresponding to the value in the final column, (MM1b) the peak intensity and beamsize.}
\tablenotetext{c}{All four components of MM1a, ($i$--$iv$), were fit simultaneously, with the position angle of ($i$) fixed to the value obtained from an initial single-component fit.}
\tablenotetext{d}{Component is assumed unresolved in the fit, and its position and flux density are fixed to the cm position and spectral index from \citet{ilee_2016}.}
\end{deluxetable*}

\section{Results}
\label{sec:results}

\subsection{1.3\,mm continuum emission}\label{sec:results_cont}
 
Figure \ref{fig:continuum}  shows two views of our new ALMA observations of G11.92$-$0.61.  Fig.~\ref{fig:continuum}a  shows a larger-scale view  ($\sim16\arcsec\times16\arcsec\sim$0.27\,pc$^{2}$), including the large-scale, well-collimated bipolar outflow from MM\,1 \citep[traced by $^{12}$CO(3--2) observed with the SMA;][]{cyganowski_2011sma}.  Fig.~\ref{fig:continuum}b shows a zoom view of the 1.3\,mm continuum emission toward MM\,1, revealing two main sources.  The dominant source, MM\,1a, is the source of the bipolar outflow (marked with a dotted line).  Situated 0\farcs57 (1920\,au) to the South-East of MM\,1a is a weaker source, MM\,1b,  which is connected to MM\,1a via smooth background emission at a level of $\sim$0.5\,mJy\,beam$^{-1}$.  Fitting in the image plane of both the compact and elongated continuum emission within $\sim$1000\,au of MM\,1a requires four individual 2D Gaussian components (see Table \ref{tab:cont}).  Peak residuals from the combination of these fits lie at the 2$\sigma$ level (0.1\,mJy\,beam$^{-1}$).  Beyond the central $\sim$1000\,au, we also report a fit to the continuum toward MM\,1b.

\subsection{CH$_{3}$CN emission}\label{sec:results_ch3cn}

\begin{figure*}
\centering
\includegraphics[width=1.0\textwidth]{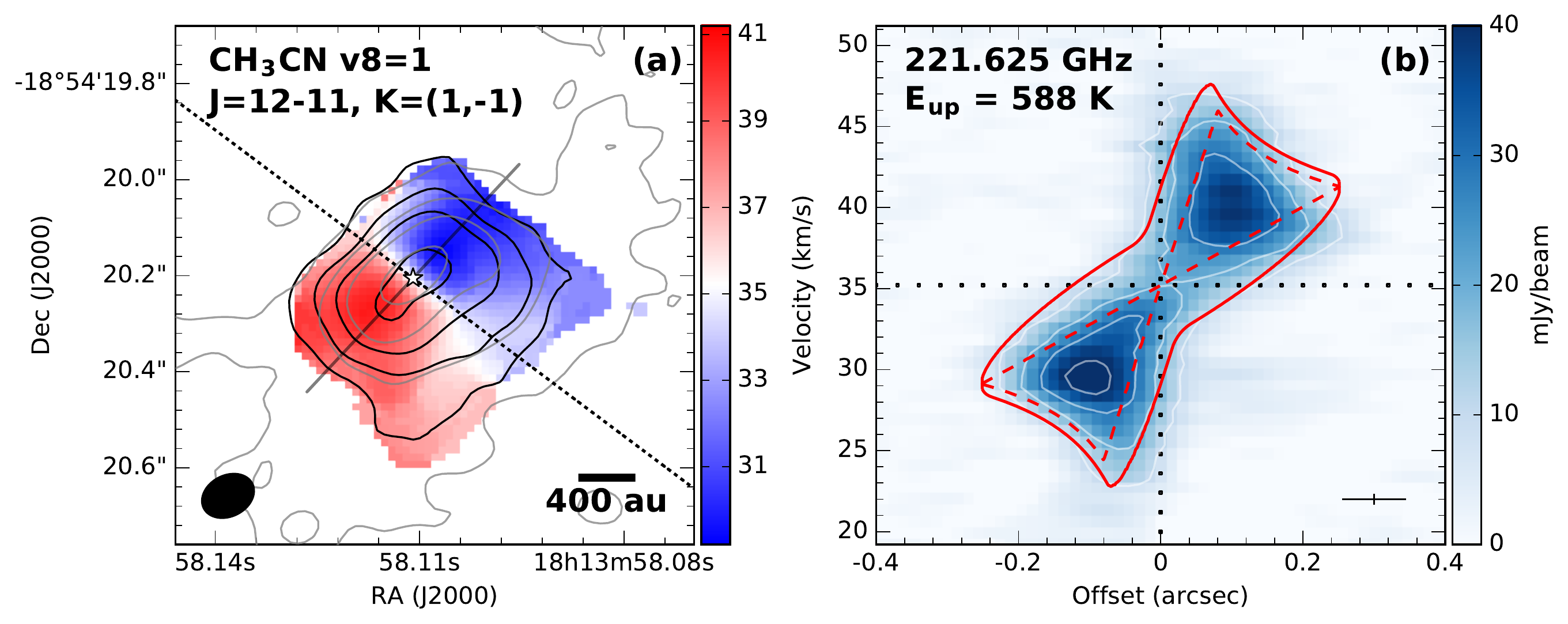}
\caption{\textbf{(a):} Integrated intensity (black contours) and intensity-weighted velocity (colorscale) of the vibrationally-excited CH$_{3}$CN $J=12$--11, $K=($1,$-$1) line ($E_{\rm up} = 588$\,K) toward MM\,1a, overlaid with the 1.3\,mm continuum contours from Figure \ref{fig:continuum} (in grey).  A white star marks the continuum peak, and the position angle of the $^{12}$CO outflow ($53^{\circ}$) is shown with a dotted line.  Integrated intensity levels: (5,15,30,60)$\sigma$,  $\sigma=6.1$\,mJy\,beam$^{-1}$\,km\,s$^{-1}$.  The beam is shown in the lower left.
\textbf{(b):} Position-velocity (PV) diagram taken along the grey line in (a).  White contours mark levels of 10, 20, 30 and 40\,mJy\,beam$^{-1}$.  Dotted black lines denote zero offset and the $v_{\rm lsr}$.  Overlaid in red are models of a thin Keplerian disk ($R_{\rm i}=270$, $R_{\rm o}=850$\,au) surrounding an enclosed mass M$_{\rm enc}$=40\,M$_{\odot}$ viewed at an inclination of 70\degrees.  The solid and dashed red lines show models with and without infalling motions at 40\% of the free-fall velocities, respectively.  The spatial and spectral resolutions are shown with a black cross.}
\label{fig:ch3cn_v8}
\end{figure*}

Figure \ref{fig:ch3cn_v8}a presents integrated intensity and intensity-weighted velocity maps of the CH$_{3}$CN $v$8=1 $K=(1,-1)$ transition (221.625\,GHz, $E_{\rm up} = 588$\,K).  The high excitation energy of this transition allows us to trace hot, dense gas within the inner 1000\,au of the circumstellar material.  The velocity field of the $v$8=1 transition exhibits rotation perpendicular to the outflow axis.  Figure \ref{fig:ch3cn_v8}b shows a position-velocity (PV) diagram for a slice along the major axis of the emission (length = 2\farcs0, PA = 129.4\degrees, centered on the continuum peak).  Both the velocity field and PV diagram are consistent with expectations for a Keplerian disk -- high central velocities showing an approximately square-root drop-off with distance.

\smallskip

Figures \ref{fig:ch3cn_k3}a \& \ref{fig:ch3cn_k3}b present integrated intensity, intensity-weighted velocity and intensity-weighted velocity dispersion maps for the CH$_{3}$CN $J=12$--11 $K=3$ transition (220.709\,GHz, $E_{\rm up} = 93$\,K).  In contrast to the $v8$=1 transition, the $K=3$ emission traces gas with a lower excitation energy and a larger spatial extent around MM\,1.  The integrated intensity map (Fig.~\ref{fig:ch3cn_k3}a) exhibits a rectangular morphology, aligned with the position angle of the bipolar outflow, which suggests the emission is tracing material in the outflow cavity. Measured opening angles from the corners of this shape are 88\degrees\ and 55\degrees\ for the North-East and South-West cavities, respectively.  The velocity field reveals a large-scale rotation pattern that is broadly consistent with the $v$8=1 transition, but with significant local deviations.  The velocity dispersion map (Fig.~\ref{fig:ch3cn_k3}b) displays a trend of increasing velocity dispersion closer to the continuum peak of MM\,1a, with additional localised increases along the outflow axis.  In particular, the area NE of MM1a shows a high dispersion, of 6-7\,km\,s$^{-1}$, which is not mirrored to the SW.  At the location of MM\,1b, deviations are observed in the integrated intensity, velocity and velocity dispersion, showing it is an outlier when compared to the surrounding material. 

\section{Discussion}
\label{sec:discussion}

\subsection{Mass estimates from the gas kinematics}\label{sec:mass_kinematics}

Using our observations of the $v$8=1 line,  we can assess the enclosed mass, $M_{\rm enc}$, within such a rotating Keplerian disk.  Following \citet{cesaroni_2011}, the expected shape of the region in PV space from which emission will originate can be expressed as
\begin{equation}
V = \sqrt{GM_{\rm enc}}\frac{x}{R^{\frac{3}{2}}} + \beta \sqrt{2GM_{\rm enc}}\frac{z}{R^{\frac{3}{2}}}, 
\end{equation}
where the first term is the  contribution from the Keplerian disk, and the second the contribution due to free fall. $V$ is  the velocity component along the line-of-sight, $M_{\rm enc}$ is the enclosed mass, $x$ and $z$  are the  co-ordinates along the disk  plane and  line-of-sight, respectively, $R = \sqrt{x^2 +  z^2}$ is the radial distance from the center of the disk (where $R_{\rm i} < R < R_{\rm o}$), and $\beta$ is a fractional factor for the contribution of the free-fall velocity. 

\smallskip

\begin{figure*}
\centering
\includegraphics[width=1.0\textwidth]{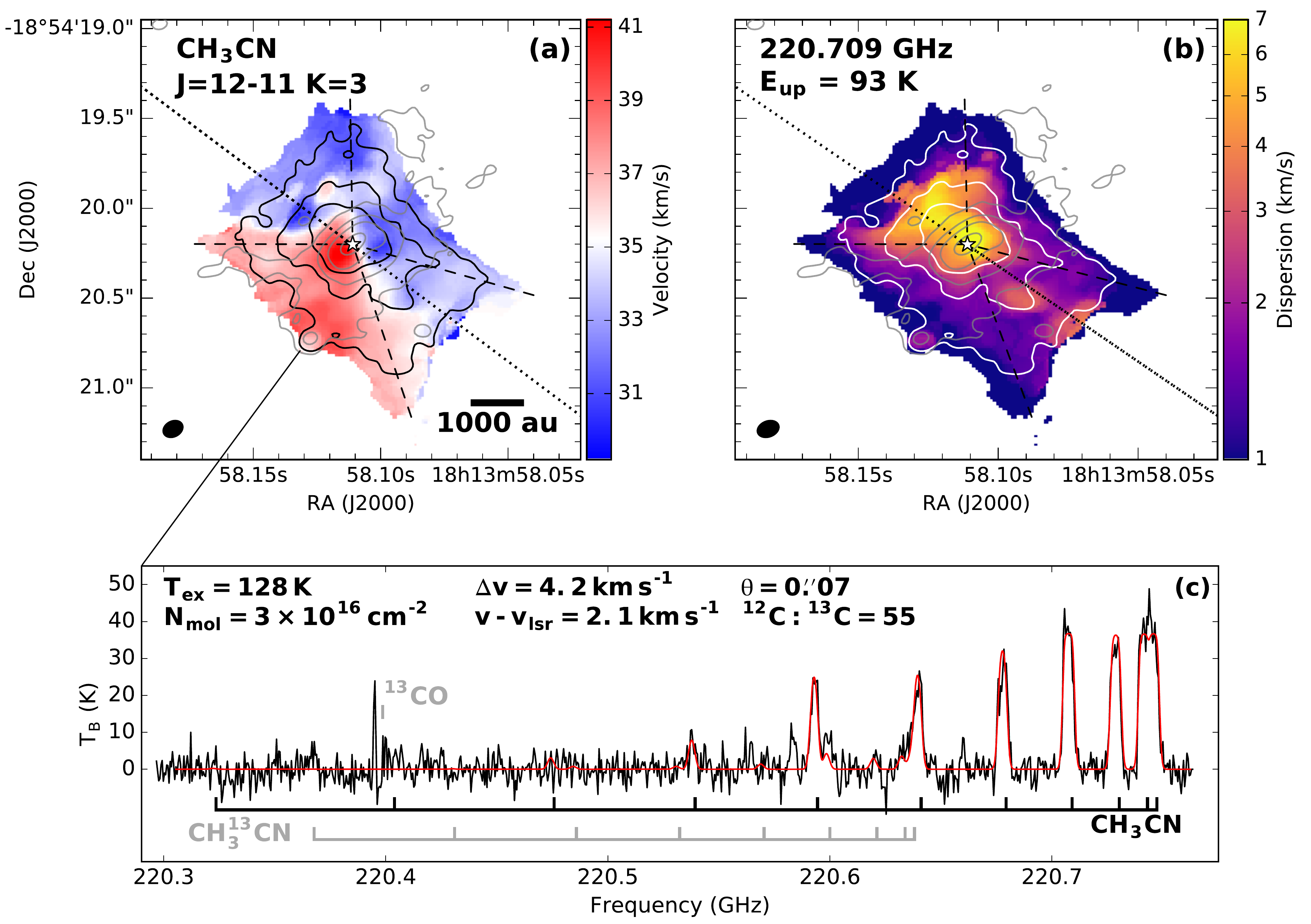}
\caption{MM1 1.3\,mm continuum contours (from Figure \ref{fig:continuum}, in grey) overlaid on \textbf{(a):} Integrated intensity (black contours) and intensity-weighted velocity (colorscale) of  the CH$_{3}$CN $J=12$--11 $K=3$ line ($E_{\rm up} = 93$\,K).  The black dotted line shows the position angle of the CO outflow, and black dashed lines the measured opening angles (NE: 88\degrees, SW: 55\degrees).  Integrated intensity levels: (5,15,30)$\sigma$,  $\sigma=11.5$\,mJy\,beam$^{-1}$\,km\,s$^{-1}$
\textbf{(b):} Integrated intensity (white contours) and intensity-weighted velocity dispersion (colorscale) of the $K=3$ line.
\textbf{(c):} The CASSIS fit (red) to the CH$_{3}$CN and CH$_{3}^{13}$CN $J = 12$--11 emission (black) at the MM\,1b continuum peak.  Best fitting parameters are labelled, and the frequencies of individual transitions are marked.
}
\label{fig:ch3cn_k3}
\end{figure*}

Our spatially resolved observations of the 1.3\,mm continuum emission (Section \ref{sec:results_cont}) allow us to break the degeneracy between an unknown disk inclination and enclosed mass.  If we assume that component ($i$) represents a flat, inclined disk, then its fitted size ($0\farcs141 \times 0\farcs050$, Table \ref{tab:cont}) corresponds to an inclination $i \sim 70$\degrees\ to the line of sight (where 0\degrees\ corresponds to a face-on disk).  Since simulations of similar disks have been shown to possess moderate aspect ratios ($H/R \lesssim 0.15$, \citealt{harries_2017}), we ascribe a conservative uncertainty of $\pm 10$\degrees\ to this inclination to account for projection effects. The inner extent of the emission is unknown, and direct measurements may be confused by significant continuum opacity \citep[see][their Section 4.3]{jankovic_2018} or chemical/radiative processes depleting the gas-phase abundance of CH$_{3}$CN.  Thus we fix the inner radius to the beamsize, $R_{\rm in} = 270$\,au.  We then perform a by-eye fit to the offset and velocity by altering the outer radius of the emission (in steps of half the geometric mean of the beamsize, $0\farcs045\sim$150\,au) and the enclosed mass (in steps of 5\,M$_{\odot}$).  Our exploration of this parameter space yields best fitting values of $R_{\rm out} = 850$\,au and $M_{\rm enc} = 40$\,M$_{\odot}$ (Figure \ref{fig:ch3cn_v8}b, dashed red line).  A purely Keplerian model does not reproduce all of the emission in PV space; to do so, our final model includes a uniform in-falling component at 40 per cent of the free-fall velocity (Figure \ref{fig:ch3cn_v8}b, solid red line).  We note that this process is unable to account for beam convolution effects, and similarly best fitting models can be obtained with $M_{\rm enc} = 40\pm5$\,M$_{\odot}$ for $i=70\mp10$\degrees.

\subsection{Physical conditions toward MM\,1b}\label{sec:cassis}

In order to determine the physical properties of the gas toward MM\,1b, we model the CH$_{3}$CN and CH$_{3}^{13}$CN emission line ladders.  Figure \ref{fig:ch3cn_k3}c shows the spectrum around the $J=12-11$ ladder extracted at the continuum peak of MM\,1b.  We utilise the CASSIS local thermodynamic equilibrium (LTE) radiative transfer package. Six free parameters were explored --- 
CH$_{3}$CN column density:   $10^{16}   < N_{\rm mol}  <  10^{19}$\,cm$^{-2}$;  
excitation temperature:  $10 <  T_{\mathrm{ex}} <  450$  K; 
line  width: $1  < \Delta  v <  10$\,km\,s$^{-1}$;  
size: $0.02\arcsec  < \theta  < 0.2\arcsec$; 
velocity:  $-5 <  v - v_{\rm lsr} < 5$\,km\,s$^{-1}$;  
isotopic ratio: $55 < \,{\mathrm{^{12}C} : \mathrm{^{13}C}} < 85$.
Fitting was performed using Markov-Chain Monte Carlo minimisation with 10$^{4}$ iterations, a cut-off parameter of 5000, and an acceptance rate of 0.5 \citep[for details see][]{ilee_2016}. The resulting best fit is shown by the red line in Figure \ref{fig:ch3cn_k3}c with parameters 
$N_{\rm mol}  =  3\times10^{16}$\,cm$^{-2}$; $T_{\mathrm{ex}} = 128$\,K; 
$\Delta  v = 4.2$\,km\,s$^{-1}$;  
$v - v_{\rm lsr} = 2.1\,$km\,s$^{-1}$;
$\theta = 0\farcs07$ and ${^{12}\mathrm{C}  : ^{13}\mathrm{C}} = 55$.

\subsection{Mass estimates from the dust emission}\label{sec:mass_dust}

Modelling of the centimeter to submillimeter wavelength SED of MM\,1 confirms that the observed  1.3\,mm  flux density is  dominated  by thermal dust emission (99.5 per cent, \citealt{ilee_2016}).  We can therefore estimate gas masses from the 1.3\,mm integrated flux densities for the various components of MM\,1.  We utilise a simple model of isothermal  dust   emission,  corrected  for  dust  opacity \citep[][Equation 3]{cyganowski_2011sma}, assuming a gas-to-dust mass ratio of 100 and a dust opacity of $\kappa_{\rm  1.3\,mm}=1.1$\,cm$^{2}$\,g$^{-1}$ (for grains with thin ice mantles and coagulation at $10^{8}$\,cm$^{-3}$; \citealt{ossenkopf_1994}).  For each component, we utilise two temperature estimates to bracket the plausible range for the circumstellar material.  In MM\,1a, for the main disk we take 150--230\,K based on the modelling of the CH$_3$CN  $J=12$--11 emission toward MM\,1 in \citet{ilee_2016}.  For the SW and W excesses, we adopt 65 -- 150\,K based on their increased radial distance from the central source.  In MM\,1b, we take 20 -- 128\,K, where the latter is based on the fit to the CH$_3$CN  $J=12$--11 ladder in Section \ref{sec:cassis}.  Under these assumptions, the mass of the main disk ranges from 0.9 -- 1.7\,M$_{\odot}$, the total mass of continuum components $i$--$iii$ ranges from 2.2 -- 5.8\,M$_{\odot}$, and the mass of MM\,1b ranges from 0.06 -- 0.6\,M$_{\odot}$.

\smallskip

Under the assumption that the gas kinematics around MM\,1b are also dominated by a Keplerian disk, we can obtain an estimate for its enclosed mass.  Using the fitted size of the 1.3\,mm continuum emission (0\farcs069, 233\,au, Table \ref{tab:cont}) and the linewidth from fits to the CH$_{3}$CN ladder (4.2\,km\,s$^{-1}$, Fig. \ref{fig:ch3cn_k3}c),  
$M_{\rm enc} = R (V / \sin i)^2 / G =  116\,{\rm au} (2.1\,{\rm km}\,{\rm s}^{-1} / \sin i)^2 / G = 0.57 / \sin^{2} i \,\mathrm{M}_{\odot}$ \citep[following][]{hunter_2014}.  Such a dynamical mass would include any contribution from a central object in MM\,1b in addition to the mass calculated from the millimeter continuum.  Therefore, the inclination of a putative disk around MM\,1b must be $\lesssim 65$\degrees\ if the central mass is $\gtrsim 0.1\,\mathrm{M}_{\odot}$.

\bigskip

\bigskip

\subsection{The MM\,1a \& b system: \\ a result of disk fragmentation?}

The combination of dynamical and continuum masses derived above allows us to place a lower limit on the mass of the central object in the MM\,1 system.  We estimate the minimum mass of the central object in MM\,1a as $(40\pm5) - 5.8 \sim 34\pm5$\,M$_{\odot}$, placing it comfortably within the O spectral class \citep{martins_2005}. In contrast, the mass derived for MM\,1b ($<$0.6\,M$_{\odot}$) corresponds to an M-dwarf or later spectral type.  The radial velocity of MM\,1b with respect to MM\,1a (2.1\kms, Section \ref{sec:cassis}) shows it is orbiting in the same sense as the Keplerian disk.  MM\,1b appears to be stable against disruption in such an orbital configuration, since the fitted size of the major axis of MM\,1b (0\farcs069, 233\,au) is comfortably within its Hill sphere:
\begin{equation}
r_{\rm H} \sim 1920\,{\rm au} \sqrt[3]{\frac{0.6 \,{\rm M}_{\odot}} {3 \times 40\,{\rm M}_{\odot}}} \sim 330\,{\rm au}.
\end{equation}

\noindent The expected orbital period of MM\,1b, $P = 1.3\times10^{4}$\,yrs, is comparable to the dynamical timescale of the bipolar outflow driven by MM\,1a ($\lesssim 10^{4}$\,yrs, \citealt{cyganowski_2011sma}).  In addition, the opening angles of the outflow cavity (88\degrees\/ and 55\degrees,  Section~\ref{sec:results_ch3cn}) are comparable to those in the simulations of \citet{kuiper_2016} at the onset of radiation pressure feedback, $<5\times10^{5}$\,yrs.  All of these observed properties point toward a young age for the MM\,1 system.

\smallskip

Our detection of MM\,1b raises the question: what is the origin of a system of objects with such an unequal mass ratio ($q \sim 0.015$)? Fragmentation of turbulent cloud cores has been shown to produce close ($\lesssim$10\,au) binary systems, but due to dynamical interactions and accretion, these binaries do not possess extreme mass ratios ($q \gtrsim 0.3$, \citealt{bate_2002}).  Fragmentation of an extended circumstellar disk is an alternative route to produce extreme mass ratio systems with larger separations \citep{clarke_2009}, which are observed on the main sequence \citep{moe_2017}.  In striking similarity to our observed properties for MM\,1a and 1b, \citet{kratter_2006} find that O stars can be expected to be surrounded by M5--G5 companions.  In addition, our measured mass for MM\,1b agrees well with predictions for masses of fragments formed via gravitational disk fragmentation at similar radii (e.g.\ 0.4\,M$_{\odot}$ at 1200\,au; \citealt{forgan_2016}).  The combined evidence thus strongly suggests that MM\,1b has formed via the fragmentation of an extended circumstellar disk around MM\,1a.

\smallskip

Finally, the fact that the central protostar powering MM1a is significantly underluminous
compared to a main sequence star of equal mass means that its energy
output is currently dominated by accretion.  
Indeed, the relative length of this evolutionary state prior
to reaching the ZAMS may determine the likelihood of formation 
of companions like MM\,1b.  This speculation can be tested by identifying
more examples of disk fragmentation around massive protostars.

\section{Conclusions}
\label{sec:conclusions}

In this Letter, we have resolved the immediate circumstellar environment of the high-mass (proto)star G11.92--0.61 MM\,1 for the first time.  Our observations show that MM\,1 separates into two main sources -- MM\,1a (the source of the bipolar outflow) and MM\,1b.  The main component of MM\,1a is an elongated millimeter-continuum structure, approximately perpendicular to the bipolar outflow.  CH$_{3}$CN $J=12-11$ $K=3$ emission traces a rotating outflow cavity, while the $v$8=1 transition exhibits a kinematic signature consistent with the rotation of a Keplerian disk. We find an enclosed mass of $40\pm5$\,M$_{\odot}$, of which 2.2--5.8\,M$_{\odot}$ can be attributed to the disk, while the mass of MM\,1b is $<0.6$\,M$_{\odot}$.  Based on the orbital properties and the extreme mass ratio of these objects, we suggest that MM\,1b is one of the first observed examples of disk fragmentation around a high mass (proto)star.  

\smallskip

Our results demonstrate that G11.92$-$0.61 MM\,1 is one of the clearest examples of a forming proto-O star discovered to date, and show its potential as a laboratory to test theories of massive (binary) star formation. 

\vspace*{0.25cm}

\acknowledgments

JDI is funded by the STFC (ST/R000549/1), and JDI and CJ~Clarke are funded by the DISCSIM project, grant agreement 341137 (ERC-2013-ADG).  CJ~Cyganowski is funded by the STFC (ST/M001296/1). DHF is funded by the ECOGAL project, grant agreement 291227 (ERC-2011-ADG).  TJ~Haworth is funded by an Imperial College Junior Research Fellowship.  TJ~Harries is funded by the STFC (ST/M00127X/1).  This paper makes use of the following ALMA data: ADS/JAO.ALMA\#2016.1.01147.S. ALMA is a partnership of ESO (representing its member states), NSF (USA) and NINS (Japan), together with NRC (Canada) and NSC and ASIAA (Taiwan) and KASI (Republic of Korea), in cooperation with the Republic of Chile. The Joint ALMA Observatory is operated by ESO, AUI/NRAO and NAOJ.  The National Radio Astronomy Observatory is a facility of the National Science Foundation operated under agreement by the Associated Universities, Inc. This research has made use of NASA's Astrophysics Data System Bibliographic Services;  Astropy,  a  community-developed core  Python  package  for Astronomy  \citep{astropy_2013};   APLpy,  an  open-source  plotting package for  Python (\url{http://aplpy.github.com}),  and the CASSIS  software  and  VADMC  databases (\url{http://www.vadmc.eu/}). CASSIS has been developed by IRAP-UPS/CNRS (\url{http://cassis.irap.omp.eu}).

%

\facilities{ALMA}


\software{ASTROPY \citep{astropy_2013}, 
          APLPY \citep{aplpy_2012},
          CASA \citep{mcmullin_2007},
          CASSIS (\url{http://cassis.irap.omp.eu}).}

\bibliographystyle{aasjournal}



\end{document}